\documentclass{amsart}

\usepackage{epsfig}
\usepackage{graphicx}
\usepackage{amsmath,amssymb,amsfonts,latexsym}
\usepackage{graphicx}

\usepackage{amssymb}
\usepackage{enumerate, xspace}
\usepackage{dsfont}
\usepackage{afterpage}
\usepackage{changebar}
\usepackage{amsthm}
\usepackage{rotating}

\numberwithin{equation}{section}

\begin{document}

\title[Second law of thermodynamics]
{Second law of thermodynamics can never be violated for all time- and space-scales}
\author {Andrew Das Arulsamy}
\address{Condensed Matter Group, Division of Interdisciplinary Science, F-02-08 Ketumbar Hill, Jalan Ketumbar, 56100 Kuala-Lumpur, Malaysia}
\email{sadwerdna@gmail.com}


\date{\today}

\begin{abstract}
Maxwell's demon was created with abilities to violate the second law of thermodynamics. But the demon always fell short of doing so because of the imposed restrictive condition that reads, the demon needs to do ``work''. Therefore, you can always preserve the second law by default. Here, we encounter the supernatural demon (without the above restriction) right at the melting point, and unambiguously expose why the demon is still doomed to fail for all time- and space-scales. This means that the second law of thermodynamics can never be violated. We develop an analytic method (based on Newton's third law) to properly analyze the entropy in a chemical system with respect to forward and reverse trajectories. We also explain the physico-chemical processes that are responsible at very short timescales. 
\end{abstract}

\maketitle

\section{Introduction}
The demon that was created by Maxwell has the ability to separate large kinetic-energy molecules from the slow-moving ones within an isolated compartment, filled with these ``fast'' and ``slow'' noninteracting molecules. The compartment is partitioned into two halves by a wall with a trap door such that the demon can open or close the door to let fast molecules (faster than the average) to go right, and the slow ones (slower or equal the average value) to go left. In doing so, the demon believed it has succeeded in decreasing the entropy, in violation of the second law of thermodynamics~\cite{max}. 

Let us assume the demon here did not do any work during the quantum information processing stage needed to acquire sufficient knowledge to decide when to open or close the trap door. This assumption is technically allowable because the demon is not part of the system. Add to that, suppose the opening and closing of this door inside the system do not require the demon to do work either. The demon is further equipped with the ability (again, without doing any work) to make its activities and itself ``invisible'' to all the observers (we do not care how all these are done) such that the observers can only detect an opening between the partitions (among all the demon's activities), and therefore detects neither the demon nor the trap door. This additional ability avoids the observer to deduce \textit{a priori} that there is ``someone'' doing work by interfering with the system. 

The preservation of the second law by invoking the condition that demands the demon to do work in its attempts to violate the second law has been addressed by others~\cite{kim3,kim4,kim5,kim6}, while interesting non-technical reviews are given in Refs.~\cite{km,jos}. Using the informational entropy initiated by Brillouin~\cite{kim4}, further advances were made leading one to assert that the foundational principle for the second law of thermodynamics is due to information loss~\cite{dun}. Here, we expose why and how the second law of thermodynamics is strictly valid by means of extensive entropy, even if the demon is equipped with some supernatural attributes stated above. The logical exposition presented here will be made unequivocal systematically such that there are no circular arguments. 

The demon proves the violation by showing that the entropy (due to disorder, $\texttt{D}$) before operating the door is 
\begin {eqnarray}
S^{\rm before}_{\texttt{D}} &=& k_{\rm B}\ln{\bigg[\frac{n^{\rm slow}_1 + n^{\rm slow}_2 + \cdots + n^{\rm fast}_1 + n^{\rm fast}_2 + \cdots}{q}\bigg]} \nonumber \\&=& k_{\rm B}\ln{\bigg[\frac{N^{\rm slow} + N^{\rm fast}}{q}\bigg]}, \label{eq:1}
\end {eqnarray}  
and after segregating the molecules, 
\begin {eqnarray}
S^{\rm after}_{\texttt{D}} &=& k_{\rm B}\ln{\bigg[\frac{n^{\rm slow}_1 + n^{\rm slow}_2 + \cdots}{q}\bigg]} + k_{\rm B}\ln{\bigg[\frac{n^{\rm fast}_1 + n^{\rm fast}_2 + \cdots}{q}\bigg]} \nonumber \\&=& k_{\rm B}\bigg[\ln{\bigg(\frac{N^{\rm slow}}{q}\bigg)} + \ln{\bigg(\frac{N^{\rm fast}}{q}\bigg)}\bigg] \label{eq:2}
\end {eqnarray}  
where $k_{\rm B}$ is the Boltzmann constant, $N^{\rm slow}$ and $N^{\rm fast}$ are the total numbers of slow and fast molecules, respectively, while $q$ is the total number of kinetic-energy states obtainable by these molecules, and $q$ is invariant such that it is independent of $N^{\rm slow}$ and $N^{\rm fast}$. In view of Eqs.~(\ref{eq:1}) and~(\ref{eq:2}), the demon concludes $S^{\rm after}_{\texttt{D}} < S^{\rm before}_{\texttt{D}}$ where $\{S^{\rm after}_{\texttt{D}},S^{\rm before}_{\texttt{D}}\} \in \mathbb{R}^-$ and $\mathbb{R}^-$ is the set of real negative numbers including zero. Here you need to be aware that $S_{\rm max} \rightarrow 0$, while $S_{\rm min} \rightarrow -\infty$ where $-\infty < 0$, which implies $S_{\rm min} < S_{\rm max}$ as it should be with negative numbers. In the subsequent sections, these inequalities will be used together with positive entropies without warning such that $S_{\rm max} \rightarrow \infty$, $S_{\rm min} \rightarrow -\infty$ and $S_{\rm min} < S_{\rm max}$, and therefore you are kindly advised to be familiar with both negative and positive entropies.

\section{Entropy production due to QPT}

Now, suppose there is a competent observer, Enthiran, a robot, who is oblivious to the demon and its activities, records the behavior, of which the fast-moving molecules accumulate in the right-hand side (r.h.s) partition, whereas the slow-moving molecules in the l.h.s partition through an opening between the two partitions. As anticipated by the demon, Enthiran writes $S^{\rm before}_{\texttt{D}}$ as an exact copy of Eq.~(\ref{eq:1}). However, the demon is thunderstruck seeing        
\begin {eqnarray}
{_{\rm after}S_{\texttt{D}}'} &=& k_{\rm B}\bigg[\ln{\bigg(\frac{N^{\rm slow}}{q}\bigg)} + \ln{\bigg(\frac{N^{\rm fast}}{q}\bigg)}\bigg] + S_{\rm QPT} \nonumber \\&\neq& S^{\rm after}_{\texttt{D}}, \label{eq:3}
\end {eqnarray}  
where $S_{\rm QPT}$ is the entropy increase due to a quantum phase transition, and of course the first term on the r.h.s of Eq.~(\ref{eq:3}) is in $\mathbb{R}^-$ while $\{S_{\rm QPT}\} \in \mathbb{R}^+$ where $\mathbb{R}^+$ is the set of real positive numbers including zero. In view of Eq.~(\ref{eq:3}), Enthiran concludes ${_{\rm after}S_{\texttt{D}}'} = S^{\rm before}_{\texttt{D}}$, claiming a type of QPT causes the respective slow and fast molecules to interact among themselves (doing work) such that $\delta W \longleftrightarrow \delta Q = T$d$S_{\rm QPT}$, contributing to an additional entropy. The robot further reasoned that ${_{\rm after}S_{\texttt{D}}'} < S^{\rm before}_{\texttt{D}}$ is not acceptable because the entropy due to an interaction (whatever it is) that causes the above QPT, logically, must be at least $S_{\rm QPT} = |S^{\rm after}_{\texttt{D}} - S^{\rm before}_{\texttt{D}}|$, if not ${_{\rm after}S_{\texttt{D}}'} > S^{\rm before}_{\texttt{D}}$. In fact, for Enthiran, the inequality, ${_{\rm after}S_{\texttt{D}}'} < S^{\rm before}_{\texttt{D}}$ is logically flawed because it certainly violates Newton's third law. \textit{Nota bene}, there is no correction to Newton's third law, in fact, both the quantum and relativity theories are required to obey Newton's third law.    

Warning---the demon did not cause the QPT to exist as supposedly observed by Enthiran, and you may correctly challenge the robot's observation saying the interaction that causes QPT does not exist in the system prepared by the demon because most of its activities are invisible to Enthiran. Indeed, and we will need your point later to show that either the demon has to do work to give rise to $S_{\rm CPT}$ such that $S_{\rm QPT} \rightarrow S_{\rm CPT} \neq 0$, or the molecules are doing the required work caused by some interactions, leading to $S_{\rm QPT} \neq 0$. Here, $S_{\rm CPT}$ is the additional entropy (replacing $S_{\rm QPT}$) due to a classical phase transition. Anyway, we have to leave the confused demon to itself in limbo (for now) because it is oblivious to quantum mechanics and (surprisingly to) Newton's third law, and prove why and how $S_{\rm QPT}$ is theoretically valid. If the molecules are non-polarizable with tightly bound outer electrons (their excitation energies are very much greater than the highest temperature of the system), and the demon did not do any work, then $S_{\rm QPT} \rightarrow S_{\rm CPT} \rightarrow 0$, and one obtains Eq.~(\ref{eq:2}). However, we will show how Maxwell's thought experiment can be captured precisely during melting of some quantum matter at the critical point, and how it satisfies the conclusion made by Enthiran, not the demon.

To technically understand why Enthiran has related the above kinetic-energy based molecular segregation procedure described by the demon to the existence of QPT, we will need to closely follow the proofs presented in Ref.~\cite{aop2}. We first melt an alkali-halide solid, LiCl. Once the melting point ($T_{\theta}$ = 610$^{\rm o}$C) is reached, we wait until approximately 50$\%$ of the solid phase are melted, and immediately isolate the system such that ``nothing'' goes in and comes out of the system. Upon observing the system at this critical point (melting point), we will find that the system needs a specific amount of thermal energy or heat ($Q_{\rm melting}$) to initiate the quantum fluctuation right at the critical point (at the melting point of LiCl) that gives rise to a wave function transformation (also known as the electronic phase transition)~\cite{ptp}. This means that any heat lower than $Q_{\rm melting}$ will not activate the wave function transformation required for the solid to liquid quantum phase transition. 

For example, any $Q > Q_{\rm melting}$ implies $Q > \langle V^{\rm e-ion}_{\rm Coulomb}\rangle + V_{\rm Waals}$ and therefore, electron-electron Coulomb repulsion ($V^{\rm e-e}_{\rm Coulomb}$) dominates, giving rise to QPT and subsequently to melting. Here $V_{\rm Waals}$ is the van der Waals attraction~\cite{prl}. In contrast, if $Q < Q_{\rm melting}$, then this simply implies $Q < \langle V^{\rm e-ion}_{\rm Coulomb}\rangle + V_{\rm Waals}$, which means the heat is insufficient to activate the repulsive $V^{\rm e-e}_{\rm Coulomb}$ required for QPT and melting. The electrons responsible for $V^{\rm e-e}_{\rm Coulomb}$ refer to the outer electrons of atomic Li and Cl. On the contrary, the electron defined by $V^{\rm e-ion}_{\rm Coulomb}$ belongs to Li, which is being attracted by Cl ion. Enough has been said. You may want to refer to Refs.~\cite{aop2,prl} for details to understand how $Q$ is related to $V^{\rm e-ion}_{\rm Coulomb}$, $V_{\rm Waals}$, entropy, specific heat, atomic energy-level spacing ($\xi$), and the quantum phase transition during melting of any molten system. If you ignore these interactions, $\langle V^{\rm e-ion}_{\rm Coulomb}\rangle$ and $V_{\rm Waals}$, then of course you can violate the second law (we defer the details to understand why this is so to another section, after the summary). 

\begin{figure}
\begin{center}
\scalebox{0.18}{\includegraphics{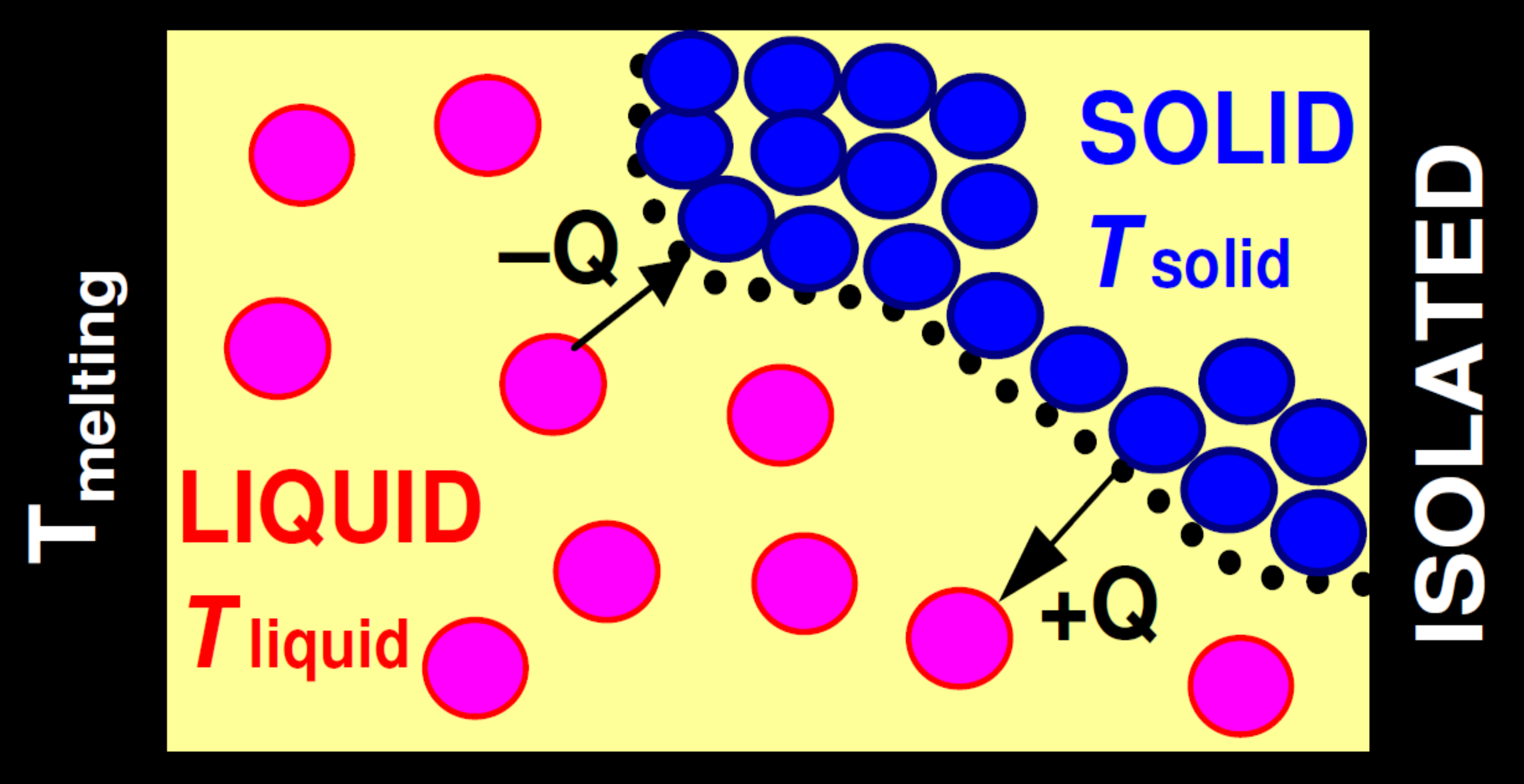}}
\caption{Solid pink circles represent both Li and Cl atoms in the liquid phase, while the solid blue circles represent the Li and Cl ions in the solid phase. The solid phase is in the top-right corner filled with solid blue circles confined by a dotted line, which also denotes the solid$|$liquid interface. The liquid-phase is distributed in the rest of the physical space. The system is isolated at the critical point, $T_{\rm melting}$ = 610$^{\rm o}$C such that the solid phase always coexists with the liquid phase where external disturbances are forbidden. The temperature distribution is such that $T_{\rm liquid} > T_{\theta} > T_{\rm solid}$ and $T_{\rm liquid} - T_{\rm solid} \ll T_{\rm melting}$. At the solid$|$liquid interface, the system is in extreme nonequilibrium due to chemical dissociations and associations where Li or Cl atom from the liquid phase may react (after releasing enough heat, $-Q$) to form permanent ionic bonds via the so-called wave function transformation, leading to quantum phase transition and solidification. On the other hand, the same amount of heat can be absorbed ($+Q$) by the Li or Cl ion in the solid phase so as to initiate the wave function transformation responsible for the ionic-bond breaking, which then allows these ions to be part of the liquid phase. Therefore, ``naturally'', the hot ions move to form the liquid phase, while the cold atoms move to be part of the solid phase, correctly replicating the system hypothesized by the Maxwell's demon. In this system, the demon does not even need to think about interfering.}
\label{fig:1}
\end{center}
\end{figure}

Anyway, Fig.~\ref{fig:1} depicts the above-stated isolated system in which, the liquid (pink solid circles) and solid (blue solid circles) phases coexist, and each phase consists of Li and Cl atoms. This system exactly replicates the demon's hypothetical system presented earlier such that the temperature of the system is ``naturally'' distributed (without any intervention from the demon) in such a way that it satisfies $k_{\rm B}T_{\rm liquid} > k_{\rm B}T_{\theta} > k_{\rm B}T_{\rm solid}$ where $k_{\rm B}T_{\rm liquid} = k_{\rm B}T_{\rm solid} + Q$, $k_{\rm B}T_{\rm liquid} - k_{\rm B}T_{\rm solid} \ll  k_{\rm B}T_{\theta}$ and $k_{\rm B}T_{\theta} = Q_{\rm melting}$. This means that the liquid is slightly hotter than the solid such that Li or Cl may release enough heat to initiate QPT and form permanent bonds with Cl or Li to form the solid phase. Conversely, Li or Cl in the solid phase can absorb enough heat to initiate QPT and to break these permanent bonds to form liquid phase. 

The above nonequilibrium processes due to heat exchanges ($\pm Q$) occur at the boundary between these two phases, which is responsible for the entropy, $S_{\rm QPT}$. The heat exchanges lead to chemical reactions (associations and dissociations) between the highly-polarized Li (with small energy-level spacing, $\xi_{\rm Li}$) and the least-polarized Cl (with large energy-level spacing, $\xi_{\rm Cl} > \xi_{\rm Li}$). For example, (Li,Cl)$^{\rm solid}_{\rm phase}$ $+ Q \rightarrow$ (Li,Cl)$^{\rm liquid}_{\rm phase}$ and (Li,Cl)$^{\rm liquid}_{\rm phase}$ $- Q \rightarrow$ (Li,Cl)$^{\rm solid}_{\rm phase}$. We can observe here that hot Li and Cl atoms (from the solid phase) move into the liquid phase, while the cold Li and Cl atoms (from the liquid phase) move back to the solid phase. Add to that, macroscopically, we may say the system is in equilibrium, but microscopically, there is an extreme nonequilibrium phenomena going on due to the above heat exchanges at the solid$|$liquid interface. 

\section{Quantum versus classical matter}

Maxwell's arguments are actually flawless and technically correct for as long as the gas molecules are noninteracting, non-polarizable classical matter, contrary to our physical universe, which is filled only with quantum matter. Maxwell's demon based its thinking on classical matter and therefore will assume that the process at the critical point (solid coexisting with the liquid) is the effect, caused by some unequal temperature distribution. The demon will falsely believe that the unequal temperature distribution is due to some unknown ``natural'' phenomena as opposed to the original system prepared by the demon earlier. For example, the demon will reason that the solid phase is in the low temperature region, while the liquid is distributed in the hotter region because the hot atoms accumulate to form the liquid phase, while the cold ones form the solid phase due to the above natural phenomena. In contrast, Enthiran will find that the chemical reactions (chemical associations and dissociations) among the atoms are the cause for the solid to coexist with its liquid, and the effect is the unequal temperature distribution stated above. 

The above summary means that Enthiran falsely reasoned the existence of a quantum phase transition that gives rise to $S_{\rm QPT}$ in the classical system prepared by the demon, while the demon falsely thought that the solid-liquid coexistence at the critical point consists of classical matter, and since the demon did not do any work, the demon falsely concludes Eq.~(\ref{eq:2}) holds, and the second law is violated ``naturally''. Obviously, we have here two ``intelligent'' beings trying to impose their respective beliefs onto each other. 

After exposing their own false arguments to them, Enthiran has had Maxwell's demon banished from our universe stating that there is only one way for the demon to prove the violation of the second law of thermodynamics--- the demon needs to find a universe filled only with classical matter and try reducing the entropy there without requiring to do sufficient work. In such a universe, quantum phase transitions will not exist, there is no quantum information processing stage, and therefore, the demon could really try to violate Newton's third law, and put its abilities of not having to do work to ``good'' use. The demon, whatever its beliefs, is doomed to fail in our universe, in one way or another because the demon either (i) needs to do significant amount of work~\cite{kim6} such that ${_{\rm after}S_{\texttt{D}}'} = S^{\rm before}_{\texttt{D}}$, or (ii) if it does not do any work, then chemical reactions due to quantum fluctuations~\cite{aop2} will be responsible for ${_{\rm after}S_{\texttt{D}}'} = S^{\rm before}_{\texttt{D}}$. 

\section{Summary}

Here, ${_{\rm after}S_{\texttt{D}}'} < S^{\rm before}_{\texttt{D}}$ has been exposed to be physico-logically invalid in our universe, filled only with quantum matter, and we arrived at this conclusion without starting with the restrictive condition that enforces the demon to do work. The only way to violate the second law in our universe is to prove $S_{\rm QPT} < |S^{\rm after}_{\texttt{D}} - S^{\rm before}_{\texttt{D}}|$ so that ${_{\rm after}S_{\texttt{D}}'} < S^{\rm before}_{\texttt{D}}$. To do that, the demon has to have the ability to fine tune the quantity $S_{\rm QPT}$ without doing sufficient work in the system. For example, recall that the demon managed not to do any work in the system it prepared earlier during opening and closing of the trap door. But the demon can never perform such a task in the system filled with quantum matter because it requires the demon to meddle with the chemical reactions (or some interactions), which occur as a result of some quantum phase transitions or electronic phase transitions or due to some wave function transformations~\cite{ptp}. For example, in the liquid phase, dynamical van der Waals ($V_{\rm Waals}$) and electron-ion Coulomb ($V^{\rm e-ion}_{\rm Coulomb}$) attractions are the dominant interactions~\cite{prl} between Li and Cl, whereas after the chemical association, one has the fluctuating bonding strength due to excited bonding electrons that could give rise to repulsive $V^{\rm e-e}_{\rm Coulomb}$ interaction between Li and Cl in the solid phase near the solid$|$liquid interface. Thus, at the solid$|$liquid interface, the wave function transformations or the electronic phase transition may take place to initiate the transition from bonding electrons (with overlapped wave functions between Li and Cl) in a solid phase, to the not-bonded electrons (with no overlapped wave functions between Li and Cl) forming the liquid phase where the not-bonded electrons are bounded strongly to their respective atoms.    

Meddling with chemical reactions means the demon needs to do work in the system to enforce the formation of solid phase from the cold atoms in the liquid phase (with large $V_{\rm Waals}$ and $V^{\rm e-ion}_{\rm Coulomb}$ attractions). Moreover, the demon also needs to do work in the system to enforce the formation of the liquid phase from the hot atoms in the solid phase that have large electron-electron (bonding electrons) repulsive interaction. Now, the above meddling requires the demon to do work ($\delta W_{\rm demon} \longleftrightarrow \delta Q_{\rm demon} = T$d$S_{\rm demon}$) such that the demon can at least partially block the chemical reactions (the above attractive and repulsive interactions) between the atoms so that $S_{\rm QPT} < |S^{\rm after}_{\texttt{D}} - S^{\rm before}_{\texttt{D}}|$ where $S_{\rm demon} \in \mathbb{R}^+$. In addition, the demon has to make sure that $S_{\rm demon} < \big|S_{\rm QPT} - (|S^{\rm after}_{\texttt{D}} - S^{\rm before}_{\texttt{D}}|)\big|$ in order to guarantee $S_{\rm QPT} < |S^{\rm after}_{\texttt{D}} - S^{\rm before}_{\texttt{D}}|$, which means the second law of thermodynamics has been violated. But this is like saying, the demon can stop a force acting in the $+x$ direction ($F^+_x$) with a smaller force opposite to $F^+_x$, in violation of Newton's third law. For example, the demon can do less work than the amount of work required to block the chemical reactions such that $S_{\rm demon} < \big|S_{\rm QPT} - (|S^{\rm after}_{\texttt{D}} - S^{\rm before}_{\texttt{D}}|)\big|$.

However, Newton's third law can never be violated, not because it has never been, but because such a violation is logically not allowed in our universe, even in the atomic-scale with extreme heat-exchange ($\pm Q$) induced quantum fluctuations. As a consequence, the demon will always end up with $S_{\rm demon} \geq \big|S_{\rm QPT} - (|S^{\rm after}_{\texttt{D}} - S^{\rm before}_{\texttt{D}}|)\big|$ due to Newton's third law. This has been correctly pointed out by the robot earlier. On the other hand, the demon is not able to justify at all, why $S_{\rm demon} < \big|S_{\rm QPT} - (|S^{\rm after}_{\texttt{D}} - S^{\rm before}_{\texttt{D}}|)\big|$ is theoretically valid, except where the demon always claims that it can do less or no work at all in quantum systems, and therefore it can violate the second law. We did let the demon off the hook all this while (since 1871) by assuming the demon can indeed do no or less work than required. However, now it is clear beyond any logical doubt that its claim is actually a mere assumption because it lacks any viable support, which is necessarily needed to justify why and how $S_{\rm demon} < \big|S_{\rm QPT} - (|S^{\rm after}_{\texttt{D}} - S^{\rm before}_{\texttt{D}}|)\big|$ is even possible logically. Consequently, it is certain now that the demon can never preserve its own assumption for as long as Newton's third law is also valid in any interacting quantum system.

\section{Entropy at short and long timescales}

From the above summary, we know the second law of thermodynamics is special because it is protected by Newton's third law such that one needs to first violate Newton's third law in order to have any chance of violating the second law of thermodynamics. Moreover, we also have discussed why the quantum phase transition, QPT (due to chemical reactions) is initiated in the presence of thermal energy (also known as heat, $Q$) if $Q > \langle V^{\rm e-ion}_{\rm Coulomb}\rangle + V_{\rm Waals}$, which then contributes to an entropy production, $S_{\rm QPT}$. However, in view of Ref.~\cite{wang}, some of you have been carried away into thinking that the isolated system presented in Fig.~\ref{fig:1}, if evaluated at short timescales, will stochastically violate the second law~\cite{sag}. Here we unequivocally show that this is not the case, and that the second law of thermodynamics can never be violated for all time- and space-scales. 

Our analyses in the subsequent paragraphs deal with a quantum phase transition at a critical point due to quantum fluctuations during a chemical reaction. Here, our strategy is opposite to Ref.~\cite{konde} in which, we define the microscopic processes that give rise to quantum-fluctuations-induced entropy production, which strictly satisfies the Newton's third law.

We recall the previously isolated molten LiCl system sketched in Fig.~\ref{fig:1}. This system is already in nanoscale, and we will consider several Li and Cl atoms in each of the two (liquid and solid) coexisting phases. The two phases coexist in an isolated system sketched in Fig.~\ref{fig:1}. A chlorine atom from the solid phase may absorb sufficient heat such that 
\begin {eqnarray}
Q > \left\langle{^{\rm attractive}}V^{\rm e(Li)-ion(Cl)}_{\rm Coulomb}\right\rangle + {^{\rm attractive}}V^{\rm Li-Cl}_{\rm Waals}, \label{eq:4}
\end {eqnarray}  
which can be used to initiate the chemical dissociation \textit{via} the quantum phase transition (due to wave function transformation). Both the Coulomb (between charges) and van der Waals (between neutral atoms) potentials in Eq.~(\ref{eq:4}) are attractive. For example, at the solid$|$liquid interface, the heat $Q$ should be large enough to overcome the Coulomb attractive force between the highly polarized (cationic forming) Li atom and the anionic forming Cl atom. In addition, this $Q$ should also be sufficient to surmount the van der Waals attractive force between the neutral Li and Cl atoms (in the absence of large polarization). 

Now, at the shortest timescale, say between $t_0 = 0$ to $t_1 < 1$ sec, sufficient amount of heat $Q$ can be absorbed by a Cl atom (in the solid phase), and be ejected out of the solid phase with a force $F^+_x$ into the liquid phase caused by the $e$-$e$ repulsive interaction ${^{\rm repulsive}_{\rm forward}}V^{\rm e(Li,Cl)-e(Cl)}_{\rm Coulomb}$. Now, it may so happen that an atomic Li (or a Cl) is on the path of this ejected Cl atom. In this case, one can indeed reverse the trajectory from $+x$ ($F^+_x$) to $-x$ ($F^-_x$), sending the Cl atom back to collide with Li or Cl in the solid phase. The probability of this reverse trajectory will be large if the number of ejected Cl or Li (from the solid phase) are much less than the number of Cl and Li in the liquid phase. Similar to a forward trajectory, the reverse trajectory is also caused by the the $e$-$e$ repulsive interaction ${^{\rm repulsive}_{\rm reverse}}V^{\rm e(Li,Cl)-e(Cl)}_{\rm Coulomb}$ where ${^{\rm repulsive}_{\rm reverse}}{\rm d}V^{\rm e(Li,Cl)-e(Cl)}_{\rm Coulomb} \longleftrightarrow \delta Q^{\rm reverse}_{\rm e-e} = k_{\rm B}T{\rm d}\Sigma^{\rm reverse}_t$ and $\Sigma^{\rm reverse}_t$ denotes the entropy associated to the reverse trajectories such that $\Sigma_t = \Sigma^{\rm forward}_t + \Sigma^{\rm reverse}_t$ (see Eq.~(\ref{eq:5})). 

Upon ejection (from the solid phase), the Cl atom may also be part of the liquid phase immediately without experiencing any reverse trajectory. It may also experience the reverse ($F^-_x$) and forward trajectories ($F^+_x$) until the atomic Cl would finally end up in the liquid or solid phase. If the probability for Cl ending up in the liquid phase is higher, then LiCl is melting away with increasing time. Now, if the above timescale or duration is short enough, say $t_1 \ll 1$ sec, then one can indeed capture the initial forward trajectory (due to ejection), followed by a reverse trajectory that sends the Cl atom back to collide with the solid phase. In the subsequent paragraphs, we will introduce and develop the proper analyses required to study the entropies due to these forward and reverse trajectories. 

The trajectory-induced dimensionless entropy ($\Sigma_t$) production for the above atomic Cl can be obtained by integrating its trajectory over a duration $t_1$, and is given by~\cite{wang}
\begin {eqnarray}
&&\Sigma_t = \frac{1}{k_{\rm B}T} \int_0^{t_1}\textbf{v}_{\rm Cl}(t)F_{\rm Cl}(t){\rm d}t, \label{eq:5}
\end {eqnarray}  
where $\textbf{v}_{\rm Cl}(t) = \dot{x}(t)$ and $F_{\rm Cl}(t) = m\ddot{x}(t)$ denote the respective velocity and force along the trajectory of the ejected Cl atom, $m$ is the mass of Cl, $x(t)$ refers to the position of the ejected Cl and $x(0)$ is the position when Cl was in the solid phase prior to ejection. Note that the duration $t_1$ depends on the type of system one studies. In our molten LiCl system, we assume $t_1 \ll 1$ sec is the duration one requires to obtain $\Sigma_t < 0$ (entropy consumption) due to $\textbf{v}_{\rm Cl} = -\textbf{v}_{\rm Cl}$. Here, the negative sign in $-\textbf{v}_{\rm Cl}$ refers to the reverse trajectory. Here, a reverse trajectory does not imply the exact opposite path of the forward trajectory~\cite{wang}. 

As expected and as it should be, if we let $t_1 \rightarrow \infty$, the reverse trajectories are negligible because they are overwhelmed by the large number of irreversible forward trajectories (meaning, LiCl is melting away) giving rise to $\Sigma_t > 0$. Now, just because you have obtained $\Sigma_t > 0$ for the limit $t_1 \rightarrow \infty$, which can be interpreted to obey the second law, you can go on and interpret the second law has been violated for $t_1 \ll 1$ sec because you have obtained $\Sigma_t < 0$. To understand why $\Sigma_t > 0$ and $\Sigma_t < 0$ are both incomplete to be exploited to determine whether they obey the second law or not, we need to first realize that the entropy is an extensive parameter, and the entropy is not a vector quantity. This (extensiveness) means that each physical process (that may occur in a system) corresponds to an entropy production, which needs to be properly taken into account. The number of physical processes and the types of processes do change with system size, and therefore, the term ``extensiveness'' is also referred to system size. Anyway, the dimensionless entropy productions for an atomic Cl due to forward and reverse trajectories, and for two different timescales ($t_1 \ll 1$ and $t'_1 \rightarrow \infty$) should be written as
\begin {eqnarray}
&&\Sigma^{\rm forward}_t = \frac{Q^{\rm forward}_{\rm e-e}}{k_{\rm B}T},~~ \Sigma^{\rm reverse}_t = \frac{Q^{\rm reverse}_{\rm e-e}}{k_{\rm B}T}, \label{eq:6}
\end {eqnarray}  
and          
\begin {eqnarray}
&&\Sigma'^{\rm ~forward}_t = \frac{Q'^{\rm ~forward}_{\rm e-e}}{k_{\rm B}T}, ~~ \Sigma'^{\rm ~reverse}_t = \frac{Q'^{\rm ~reverse}_{\rm e-e}}{k_{\rm B}T}, \label{eq:7}
\end {eqnarray}  
respectively, where 
\begin {eqnarray}
&&\Sigma^{\rm forward}_t + \Sigma^{\rm reverse}_t = \Sigma_t = \Sigma_t(t_1 \ll 1), \label{eq:77a} \\&& 
\Sigma'^{\rm ~forward}_t + \Sigma'^{\rm ~reverse}_t = \Sigma'_t = \Sigma_t(t'_1 \rightarrow \infty), \label{eq:77b}
\end {eqnarray}  
and $\Sigma_t$ is given by Eq.~(\ref{eq:5}). Note that we have dropped the unit for time ``sec'' from $t_1 \ll 1$ sec for convenience and also because it is arbitrary (system dependent). The superscripts, ``forward'' and ``reverse'' refer to forward and reverse trajectories, respectively. The forward trajectory is initiated by the Coulomb repulsive interaction (${^{\rm repulsive}_{\rm forward}}V^{\rm e(Li,Cl)-e(Cl)}_{\rm Coulomb}$) giving rise to 
\begin {eqnarray}
&&\frac{{\rm d}S^{\rm forward}_{\rm e-e}}{k_{\rm B}} = \frac{\delta Q^{\rm forward}_{\rm e-e}}{k_{\rm B}T} \longleftrightarrow (k_{\rm B}T)^{-1}{^{\rm repulsive}_{\rm forward}}{\rm d}V^{\rm e(Li,Cl)-e(Cl)}_{\rm Coulomb}, \label{eq:7a}
\end {eqnarray}  
For the reverse trajectory, one has 
\begin {eqnarray}
&&\frac{{\rm d}S^{\rm reverse}_{\rm e-e}}{k_{\rm B}} = \frac{\delta Q^{\rm reverse}_{\rm e-e}}{k_{\rm B}T} \longleftrightarrow (k_{\rm B}T)^{-1}{^{\rm repulsive}_{\rm reverse}}{\rm d}V^{\rm e(Li,Cl)-e(Cl)}_{\rm Coulomb}. \label{eq:7b}
\end {eqnarray}  
Equations~(\ref{eq:7a}) and~(\ref{eq:7b}) mean that 
\begin {eqnarray}
&&{^{\rm repulsive}_{\rm forward}}{\rm d}V^{\rm e(Li,Cl)-e(Cl)}_{\rm Coulomb} \longleftrightarrow \delta Q^{\rm forward}_{\rm e-e} = k_{\rm B}T{\rm d}\Sigma^{\rm forward}_t, \label{eq:8} \\&& 
{^{\rm repulsive}_{\rm reverse}}{\rm d}V^{\rm e(Li,Cl)-e(Cl)}_{\rm Coulomb} \longleftrightarrow \delta Q^{\rm reverse}_{\rm e-e} = k_{\rm B}T{\rm d}\Sigma^{\rm reverse}_t. \label{eq:8a} 
\end {eqnarray}
Equations~(\ref{eq:8}) and~(\ref{eq:8a}) were used to arrive at Eqs.~(\ref{eq:6}) and~(\ref{eq:7}). Subsequently, we use Eqs.~(\ref{eq:6}) and~(\ref{eq:7}) to obtain 
\begin {eqnarray}
&&\Sigma^{\rm forward}_t - |\Sigma^{\rm reverse}_t| = \frac{Q^{\rm forward}_{\rm e-e}}{k_{\rm B}T} - \frac{Q^{\rm reverse}_{\rm e-e}}{k_{\rm B}T}, \label{eq:9} \\&& \Rightarrow {\rm d}\Sigma_t = \frac{\delta Q_{\rm e-e}}{k_{\rm B}T}, \label{eq:10} \\&& 
\Sigma'^{\rm ~forward}_t - |\Sigma'^{\rm ~reverse}_t| = \frac{Q'^{\rm ~forward}_{\rm e-e}}{k_{\rm B}T} - \frac{Q'^{\rm ~reverse}_{\rm e-e}}{k_{\rm B}T}, \label{eq:11} \\&& \Rightarrow {\rm d}\Sigma'_t = \frac{\delta Q'_{\rm e-e}}{k_{\rm B}T}. \label{eq:12} 
\end {eqnarray}
Here, we have noted that unlike a force ($F(t)$) or a velocity ($\textbf{v}(t)$), entropy is not a vector quantity, and therefore, we have enforced positivity by writing $|\Sigma^{\rm reverse}_t|$ and $|\Sigma'^{\rm ~reverse}_t|$. \textit{Nota bene}, you cannot assign a ``$+$'' or a ``$-$'' sign to an entropy (a scalar quantity) just because you have determined the entropy from a forward or a reverse trajectory (a vector quantity). Such an assignment is entirely \textit{ad hoc} and arbitrary. This is the first reason why positive entropy production determined from the above trajectories is not provable. The second reason is that not all physical processes (giving rise to entropy production) can be captured solely with $\Sigma_t$ (see Eqs.~(\ref{eq:13}) and~(\ref{eq:14})) because not all physical processes translate into forward and/or reverse trajectories. 

Equations~(\ref{eq:10}) and~(\ref{eq:12}) are for reversible processes where the total entropy remains the same (does not increase or decrease). For irreversible processes, Eqs~(\ref{eq:10}) and~(\ref{eq:12}) read 
\begin {eqnarray}
&&{\rm d}\Sigma_t < \frac{\delta Q_{\rm e-e}}{k_{\rm B}T} + {\rm d}S_{\rm disorder} + {\rm d}S^{\rm other}_{\rm process}, \label{eq:13} \\&& {\rm d}\Sigma'_t < \frac{\delta Q'_{\rm e-e}}{k_{\rm B}T} + {\rm d}S'_{\rm disorder} + {\rm d}S'^{\rm ~other}_{\rm process}, \label{eq:14} 
\end {eqnarray}
respectively. In Eq.~(\ref{eq:13}) however, we can correctly assume $S_{\rm disorder} = 0 = S^{\rm other}_{\rm process}$ because the timescale $t_1 \ll 1$ is very short such that Eq.~(\ref{eq:13}) can be replaced by Eq.~(\ref{eq:10}) because at the shortest timescale, the process is reversible at worst. Apart from that, $S^{\rm other}_{\rm process}$ is the collection of entropies due to other physical processes that occur in a given system, which we are not aware of. One should note here that melting is a irreversible process when the timescale is not small or when $t'_1 \rightarrow \infty$. In this case, it is never possible to reverse all the trajectories and all the processes that have occurred during the forward trajectories, without doing sufficient work. To reverse these forward trajectories (following the exact paths), we will always end up doing more work than the original amount of work required for forward trajectories, thus giving rise to an inevitable entropy production. You may want to recall why and how the Newton's third law and the quantum phase transitions (both were explained earlier) made sure of the above never-possible scenario.    
Apparently, in view of Eqs.~(\ref{eq:13}) and~(\ref{eq:14}), we can violate the second law if
\begin {eqnarray}
&&\Sigma^{\rm forward}_t - |\Sigma^{\rm reverse}_t| > \frac{Q^{\rm forward}_{\rm e-e}}{k_{\rm B}T} - \frac{Q^{\rm reverse}_{\rm e-e}}{k_{\rm B}T}, \label{eq:15} ~~{\rm or} \\&& \Sigma'^{\rm ~forward}_t - |\Sigma'^{\rm ~reverse}_t| > \frac{Q'^{\rm ~forward}_{\rm e-e}}{k_{\rm B}T} - \frac{Q'^{\rm ~reverse}_{\rm e-e}}{k_{\rm B}T}. \label{eq:16} 
\end {eqnarray}
We focus on Eq.~(\ref{eq:15}) because the inequality in Eq.~(\ref{eq:16}) is never possible due to Eq.~(\ref{eq:14}). For short timescales ($t_1 \ll 1$) following Eq.~(\ref{eq:15}), $Q^{\rm forward}_{\rm e-e}(k_{\rm B}T)^{-1}$ and $Q^{\rm reverse}_{\rm e-e}(k_{\rm B}T)^{-1}$ given in Eqs.~(\ref{eq:15}) and~(\ref{eq:16}) are the sources that have been completely translated into forward and reverse trajectories, which then give rise to $\Sigma^{\rm forward}_t$ and $\Sigma^{\rm reverse}_t$, respectively. Therefore, it is immediately obvious that the inequality in Eq.~(\ref{eq:15}) or in Eq.~(\ref{eq:16}) is in violation of Newton's third law. 

Anyway, all we have to do now is to ignore the source that gives rise to $\Sigma^{\rm forward}_t$ by taking $Q^{\rm forward}_{\rm e-e}(k_{\rm B}T)^{-1} = 0$. This can be done by choosing $t_x$ such that the repulsive interaction between a Cl and a Li or between a Cl and another Cl is not recorded. This means that $t_x$ starts the counting after the repulsive interaction (that gives rise to forward trajectory), and the counting stops at $t_y$ when the Cl atom (which is now in reverse trajectory) comes to a halt. Indeed, the duration $t_x - t_y$ is very small such that $t_x - t_y = t_1 \ll 1$. Consequently, and after noting Eq.~(\ref{eq:6}), Eq.~(\ref{eq:15}) reads   
\begin {eqnarray}
&&\Sigma^{\rm forward}_t - |\Sigma^{\rm reverse}_t| > 0 - \frac{Q^{\rm reverse}_{\rm e-e}}{k_{\rm B}T}. \label{eq:17} 
\end {eqnarray}
Here, Eq.~(\ref{eq:17}) is true regardless whether $\Sigma^{\rm forward}_t > |\Sigma^{\rm reverse}_t|$ or $\Sigma^{\rm forward}_t < |\Sigma^{\rm reverse}_t|$, and therefore, the violation is guaranteed as it should be because we did not treat the entropy as a vector quantity. We are basically done explaining why and how the entropy can be violated for short timescales ($t_x - t_y \ll 1$). The violation of the second law is possible if we ignore a certain physico-chemical process, and in this case, the $e$-$e$ repulsive interaction that gives rise to forward trajectories has been ignored. Having said that, it should be clear now why such a violation was actually never observed in Ref.~\cite{wang}.

Following Ref.~\cite{wang}, if you have a colloidal particle that is trapped by means of a laser-induced force in the background of moving water (H$_2$O)~\cite{wang}, then Eq.~(\ref{eq:15}) representing this colloidal particle-water system should read 
\begin {eqnarray}
\Sigma^{\rm forward}_t - |\Sigma^{\rm reverse}_t| &=& \frac{Q^{\rm Hyd-B}_{\rm break}}{k_{\rm B}T} - \frac{Q^{\rm Hyd-B}_{\rm form}}{k_{\rm B}T}, \label{eq:18}
\end {eqnarray}
where $Q^{\rm Hyd-B}_{\rm break}$ and $Q^{\rm Hyd-B}_{\rm form}$ are the thermal energies due to hydrogen-bond breaking and forming, respectively, giving rise to the respective forward and reverse trajectories, which are then translated into their respective entropies $\Sigma^{\rm forward}_t$ and $|\Sigma^{\rm reverse}_t|$. Here, water molecules are quantum matter that can form non-permanent hydrogen bonds (Hyd-B) with other water molecules in the liquid phase. This means that the hydrogen bonds are being broken and formed constantly and randomly due to fluctuating $e$-$e$ interaction in the liquid phase~\cite{her1,her2,her3,pccp}. On the contrary, the water molecules in ice have permanent hydrogen bonds.      

In the system prepared by Wang \textit{et al}.~\cite{wang}, the force provided by the laser on the colloidal particle activates $Q^{\rm Hyd-B}_{\rm break}$ that initiates the forward trajectory. Here, we also assumed that there are no other microscopic processes involved such as the micro-turbulence when the colloidal particle is trapped in the background of slow-moving water and the laser-water molecule interaction~\cite{wang}. The existence of any other processes will only strengthen the validity of the second law (see Eq.~(\ref{eq:13})). On the other hand, $Q^{\rm Hyd-B}_{\rm form}$ activates the reverse trajectory, and subsequently, giving rise to its associated entropy $|\Sigma^{\rm reverse}_t|$. When $t_1 \ll 1$, the inequality $\Sigma^{\rm forward}_t < |\Sigma^{\rm reverse}_t|$ is due to $Q^{\rm Hyd-B}_{\rm break} < Q^{\rm Hyd-B}_{\rm form}$. Therefore, the observation made in Ref.~\cite{wang} unambiguously proves that their system is reversible (see Eq.~(\ref{eq:18})) for short timescales, and there is absolutely no violation of the second law of thermodynamics.         

\section{Conclusions}

We have logically shown that the original Maxwell's arguments are actually flawless and technically correct, if and only if the system consists of classical matter. We also have shown that the second law of thermodynamics can never be violated in any quantum matter because it is protected by Newton's third law, even down to an atomic- and/or nano-scale due to the existence of quantum phase transitions. The above conclusion is based on the analyses carried out on a LiCl molten system--- this system is an exact replication of Maxwell's thought experiment, which has been isolated at the melting point. Apart from that, we also have developed an analytic technique to properly evaluate the entropies arising from the forward and reverse trajectories, without falsely interpreting the entropy as a vector quantity. Using this technique, we have unambiguously proven that the second law of thermodynamics is inviolable for all time- and space-scales supported by the experimental results reported by Wang \textit{et al}.~\cite{wang}. All our analyses are consistent with well-established logical foundation (inequalities, positive and negative numbers), basic physics ($e$-$e$ Coulomb interaction, Newton's third law, quantum phase transitions, scalar and vector quantities) and chemistry (LiCl melting and fluctuating hydrogen bonds in water).         

\section*{Acknowledgments}
This work was supported by Sebastiammal Innasimuthu, Arulsamy Innasimuthu, Amelia Das Anthony, Malcolm Anandraj and Kingston Kisshenraj. Special thanks to Mir Massoud Aghili Yajadda (CSIRO, Lindfield) for providing most of the listed references and also to Eugene Tam (CSIRO, Melbourne) for the prompt delivery of Ref.~\cite{wang}. I would like to thank the referee for the suggestions leading me to Ref.~\cite{wang}.

\end{document}